\documentclass[aps,pra,preprint,superscriptaddress,showpacs]{revtex4}

\usepackage[dvips]{graphicx}

\begin{document}


\title{Fourier Synthesis of Optical Potentials for Atomic Quantum 
Gases}


\author{Gunnar Ritt}
\affiliation{Physikalisches Institut der Universit\"at T\"ubingen,
Auf der Morgenstelle 14, 72076 T\"ubingen, Germany}

\author{Carsten Geckeler}
\affiliation{Physikalisches Institut der Universit\"at T\"ubingen,
Auf der Morgenstelle 14, 72076 T\"ubingen, Germany}

\author{Tobias Salger}
\affiliation{Physikalisches Institut der Universit\"at T\"ubingen,
Auf der Morgenstelle 14, 72076 T\"ubingen, Germany}
\affiliation{Institut f\"ur Angewandte Physik der Universit\"at Bonn,
Wegelerstra\ss e 8, 53115 Bonn, Germany}

\author{Giovanni Cennini}
\affiliation{Physikalisches Institut der Universit\"at T\"ubingen,
Auf der Morgenstelle 14, 72076 T\"ubingen, Germany}

\author{Martin Weitz}
\affiliation{Physikalisches Institut der Universit\"at T\"ubingen,
Auf der Morgenstelle 14, 72076 T\"ubingen, Germany}
\affiliation{Institut f\"ur Angewandte Physik der Universit\"at Bonn,
Wegelerstra\ss e 8, 53115 Bonn, Germany}


\date{\today}

\begin{abstract}
We demonstrate a scheme for the Fourier synthesis of periodic optical 
potentials with asymmetric unit cells for atoms. In a proof of 
principle experiment, an atomic Bose-Einstein condensate is exposed 
to either symmetric or sawtooth-like asymmetric potentials by 
superimposing a conventional standing wave potential of $\lambda/2$ 
spatial periodicity with a fourth-order lattice potential of 
$\lambda/4$ periodicity. The high periodicity lattice is realized 
using dispersive properties of multiphoton Raman transitions. Future 
applications of the demonstrated scheme could range from the search 
for novel quantum phases in unconventionally shaped lattice 
potentials up to dissipationless atomic quantum ratchets.
\end{abstract}


\maketitle

\newcommand{\vek}[1]{{\bf #1}}

The properties of solid state materials are much determined by the 
internal spatial structure of its constituents. Most natural solid 
state materials are crystals, in which the constituents are spatially 
ordered following the principle of lattice structures. A large class 
of different unit cells is known, whose classification into crystal 
classes, point- and space groups is well established within the field 
of crystallography \cite{hahn}. Recently, artificial crystals of 
atoms bound by light, so called optical lattices, have developed into 
model systems for solid state physics problems. In such systems, 
effects like Bloch-oscillations and the Mott-insulator transition 
were observed \cite{dahan,greiner}. On the other hand, current 
quantum gas experiments with optical lattices have been limited to 
sinusoidal atom potentials. Recent theoretical work pointed out that 
in unconventional lattice structures novel quantum phases are 
expected \cite{buonsante,santos}. A possible way to at least in one 
dimension produce arbitrarily shaped potentials is the use of beams 
that are inclined under different angles, which may seem impractical 
and hard to phase stabilize especially if larger numbers of beams are 
required. Notably, such phase stabilization has recently been 
achieved for two near-resonant standing waves \cite{stuetzle}. For 
thermal atoms some nonstandard optical lattices, as subwavelength 
structure potentials or asymmetric lattices, have been realized with 
magneto-optical potentials or grey lattices 
\cite{pfau,mennerat-robilliard}. These schemes require near-resonant 
optical radiation whose inherent dissipation impedes the use of 
quantum degenerate atomic gases. With the so far demonstrated 
schemes, the class of realizable lattice potentials furthermore seems 
limited.

Here, we demonstrate a scheme to create dissipationless optical 
lattice potentials with $\lambda/4$ spatial periodicity, which is 
realized with a fourth-order Raman process. Building upon this 
potential, a Fourier synthesis of lattice potentials is demonstrated. 
By combining the $\lambda/4$ period multiphoton potential with a 
conventional lattice of $\lambda/2$ spatial periodicity with 
appropriate phase, an asymmetric lattice with variable asymmetry is 
realized. The scheme for the generation of the $\lambda/4$ period 
lattice potential is generalizable, in principle, to optical lattice 
potentials with $\lambda/2n$ spatial periodicity, where $n$ is an 
integer number. By diffraction of a rubidium Bose-Einstein condensate 
at the optical light shift potential, absorption images of atom 
clouds closely connected to the reciprocal lattice of the synthesized 
potential structures were observed.

In conventional optical lattices with two counterpropagating beams of 
wavelength $\lambda$, the resulting optical potential $V(z) = 
-(\alpha/2)|E(z)|^2$, with $\alpha$ as the atomic polarizability, is 
proportional to $\cos^2 kz = (1+\cos 2kz)/2$, yielding the well-known 
$\lambda/2$ spatial periodicity. The atoms here undergo virtual 
two-photon processes of absorption of a photon from one laser beam 
and stimulated emission into the counterpropagating beam. For a 
Fourier synthesis of arbitrarily shaped potentials, the required 
higher harmonics can, in principle, be generated by using a 
combination of standing waves with wavelengths of the fractional 
harmonics $\lambda/n$, where $n$ is integer, yielding a lattice 
potential of spatial period $\lambda/2n$. A lattice potential of the 
same periodicity could also be produced, in case each virtual 
absorption or stimulated cycle could be replaced by a simultaneous 
process induced by $n$ photons of the fundamental wavelength 
$\lambda$ absorbed or emitted along the corresponding direction (see 
Fig.~1a). The spatial period of this higher order lattice potential 
can be expressed as $\lambda_{\mathrm{eff,}n}/2 = \lambda/2n$, where 
$\lambda_{\mathrm{eff,}n} = \lambda/n$ denotes the effective 
wavelength of a $n$-photon field. A direct measurement of this 
effective multiphoton wavelength is in general a very difficult task, 
as has been discussed in the context of Heisenberg-limited 
measurements \cite{jacobson}. We here use the high frequency 
resolution of Raman transitions between ground state sublevels to 
separate in frequency space the desired $2n$-th order process from 
lower order contributions. Theoretical work has predicted, that a 
lattice potential of $\lambda/2n$ spatial periodicity is indeed 
expected with this approach \cite{berman,dubetsky,weitz}. For early 
theoretical work analyzing the use of Raman transitions to create 
optical potentials for atoms, see \cite{hope}. For experimental work 
on an improved spectral resolution with multiphoton Raman 
transitions, see \cite{cataliotti}. Other work has demonstrated, that 
a near-resonant Raman lattice can cool atoms to a few microkelvins 
\cite{zhang}. 

Fig.~1b shows the used scheme for a four-photon 
lattice, which yields a periodic potential with $\lambda/4$ spatial 
periodicity. The scheme uses three-level atoms with two stable ground 
states $|\mathrm{g_0}\rangle$ and $|\mathrm{g_1}\rangle$ and one 
electronically excited level $|\mathrm{e}\rangle$. Compared to the 
four-photon ladder scheme (right graph of Fig.~1a), one absorption 
and one stimulated emission process have been interchanged by a 
stimulated emission and absorption process of an oppositely directed 
photon respectively. To avoid second order standing wave processes, a 
minimum of three laser frequencies is required. The atoms are 
irradiated with two driving optical fields of frequencies 
$\omega\pm\Delta\omega$ from the left and one field with frequency 
$\omega$ from the right. A non-zero two-photon detuning $\delta$ is 
used to suppress resonant Raman processes. The adiabatic light shift 
potential for e.\,g.\ atoms in state $|\mathrm{g_0}\rangle$ here 
exhibits the desired fourth-order energy shift $V \propto (1+\cos 
4kz)$ with spatial periodicity $\lambda/4$ 
\cite{berman,dubetsky,weitz}. The depth of the optical potential is 
given by $(\hbar/2\delta) \cdot \Omega_\mathrm{eff}^\mathrm{+} 
\Omega_\mathrm{eff}^\mathrm{-}$ with $\Omega_\mathrm{eff}^\mathrm{+} = 
\Omega_\mathrm{g0,e}^\mathrm{+}\Omega_\mathrm{e,g1}^\mathrm{0}/2\Delta$ and 
$\Omega_\mathrm{eff}^\mathrm{-} = 
\Omega_\mathrm{g1,e}^\mathrm{-}\Omega_\mathrm{e,g0}^\mathrm{0}/2\Delta$. 
Here, 
$\Omega_\mathrm{g0,e}^\mathrm{+}$ and $\Omega_\mathrm{e,g1}^\mathrm{0}$  
($\Omega_\mathrm{g0,e}^\mathrm{0}$ and $\Omega_\mathrm{e,g1}^\mathrm{-}$)
are the Rabi frequencies for the transitions $|\mathrm{g_0}\rangle 
\Leftrightarrow |\mathrm{e}\rangle$ and $|\mathrm{e}\rangle 
\Leftrightarrow |\mathrm{g_1}\rangle$
driven by the light fields with frequency 
$\omega+\Delta\omega$ and $\omega$ ($\omega$ and $\omega-\Delta\omega$)
respectively. Further, $\Delta$ denotes the 
optical detuning from the excited atomic level 
$|\mathrm{e}\rangle$, which with $\Delta \gg \Delta\omega$ is
here assumed to be identical for all three optical frequencies. 
It should be pointed out that the four-photon energy shift is 
not proportional to $I^2$, i.\,e.\ the effect does not originate from 
the nonlinearity of the multiphoton transitions \cite{bentley}. In an 
atom optics picture, the obtained $\lambda/4$ spatial periodicity can 
be understood by considering the momentum transfer to an atom during 
virtual four-photon cycles of alternating absorption and stimulated 
emission which is $\pm 4\hbar k$, being a factor 2 above the 
corresponding processes in a standing laser wave. By combining 
lattice potentials of different spatial periodicities, in principle, 
arbitrarily shaped periodic potentials can be generated.

In our experimental setup (see Fig.~2), light detuned some 2\,nm to 
the red of the rubidium D2-line is generated by a high-power diode 
laser system. The emitted radiation is split into two beams (beam\,1, 
beam\,2), from which, after passing independent aousto-optical 
modulators (AOM\,1, AOM\,2), the two counterpropagating beams 
realizing the lattice potentials are derived. The AOMs are used for 
switching of the optical fields and in the case of beam\,2, to 
superimpose several optical frequency components onto a single beam 
path. To generate the rf drive signals for the AOMs, we use a set of 
four phase-locked function generators. While AOM\,1 is driven with a 
single radio frequency, for AOM\,2, the radio frequency signals of 
three function generators are combined. In this way, the three 
different optical frequencies required to synthesize the asymmetric 
lattice can be generated in beam\,2. After passing the AOMs, the two 
beams are fed through optical fibers and focused in a 
counterpropagating geometry onto a Bose-Einstein condensate (BEC) 
confined in a CO$_2$-laser dipole trap. The typically used optical 
lattice beam powers are 18\,mW for the frequency components 
$\omega\pm\Delta\omega$, and 24\,mW (4\,mW) for the components with 
frequency $\omega$ in beam\,1 (beam\,2) on a 40\,$\mu$m beam waist. 
The optical lattice beams are inclined under an angle of 41$^\circ$ 
respectively to the horizontally orientated CO$_2$-laser trapping 
beam.

Our $^{87}$Rb BEC is produced all-optically by evaporative cooling in 
a tightly focused CO$_2$-laser beam operating near a wavelength of 
10.6\,$\mu$m, as described in detail in \cite{cennini}. By activating 
a magnetic gradient field during the evaporation phase, a 
spin-polarized condensate with approximately 10,000 atoms in the 
$m_F=-1$ Zeeman state of the $F=1$ ground state is produced. A 
magnetic bias field generates a frequency splitting 
$\omega_\mathrm{Z}/2\pi \approx 805$\,kHz between neighboring Zeeman 
ground states of the rubidium atoms. For a typical optical frequency 
difference $\Delta\omega/2\pi = 930$\,kHz we arrive at a Raman 
detuning $\delta/2\pi \approx 125$\,kHz. The magnetic bias field 
forms an angle with respect to the optical beam axis, so that the 
atoms experience both $\sigma^+$, $\sigma^-$, as well as $\pi$ 
polarizations. To realize the multiphoton lattice as shown in 
Fig.~1b, we use the $F=1$ ground state levels $m_F=-1$ and $m_F=0$ as 
states $|\mathrm{g_0}\rangle$ and $|\mathrm{g_1}\rangle$ respectively 
and the $5\mathrm{P}_{3/2}$ manifold as the electronically excited 
state $|\mathrm{e}\rangle$. Since the lattice beam detuning is orders 
of magnitude larger than the upper state hyperfine structure 
splitting, Raman transitions only couple neighboring ground state 
Zeeman levels. For alkali atoms, the Raman lattice detuning is 
limited to a value of order of the excited state fine structure 
splitting ($\approx {}$15\,nm for Rb, 34\,nm for Cs), which for these 
atoms for typical parameters limits the expected coherence time of 
the multiphoton lattice to a value of order of a second due to photon 
scattering. Longer coherence times should be possible e.\,g.\ with 
the use of metastable P-triplett states of two-electron atoms, as 
e.\,g.\ the $^3\mathrm{P}_2$ state in atomic ytterbium with 17\,s 
life time, an atom with which BEC has recently been achieved 
\cite{takasu}. After the generation of the BEC, the lattice beams are 
activated for a period $T = {}$\,6$\mu$s. To read out the spatially 
varying potential of the multiphoton lattice, we study the far-field 
diffraction pattern of the imprinted phase grating after a free 
falling time of 10~ms. From the measured diffraction patterns, we can 
reconstruct the optical potential.

As is well known from crystal structure analysis, measurements of the 
intensity of a scattered or diffracted beam, do not in general allow 
one to reconstruct the lattice structure, as is known as the phase 
problem. For many special cases, a reconstruction is however possible 
\cite{bergmann}. In our experiment, assuming that the one-dimensional 
lattice is invariant to spatial translations of $\lambda/2$, the 
reciprocal lattice is limited to the harmonics $\lambda/2n$ and it is 
straightforward to show that a measurement of the atom diffraction 
pattern $W(p) = |\Psi(p)|^2$ allows one to reconstruct the spatial 
lattice structure.

We have fitted the result of a numerical solution of the 
time-dependent, momentum-picture Schr\"odinger equation to our 
experimental far-field diffraction data. The used Hamiltonian was $H 
= p^2/2m+V(z)$, where $V(z) = 
\frac{V_1}{2}\cos(2kz)+\frac{V_2}{2}\cos(4kz+\Phi)$. For known phase 
$\Phi$ and pulse length $T$, the potential depths $V_1$ and $V_2$ of 
the two- and four-photon lattice components were left as free 
parameters and could thus be determined from the fit. Interestingly, 
we found no major difference between these results and an analytical 
solution for the atomic diffraction pattern obtained using the simple 
thin grating (Raman-Nath) approximation (roughly below 20\% 
difference). In the Raman-Nath approximation, the kinetic energy in 
the Hamiltonian is neglected, which requires (1) $T \ll 
1/\omega_\mathrm{rec}$, where $\omega_\mathrm{rec}$ is the recoil 
frequency, and (2) $T \ll T_\mathrm{osc}/4$, where $T_\mathrm{osc}$ 
is the oscillation period \cite{gupta}. The first condition is 
fulfilled in our experiment, whereas the second is not. The 
difference between the numerical solution and the predictions in the 
Raman-Nath approximation is due to the neglect of the center-of-mass 
motion in this approximation.

Within the Raman-Nath regime, a quite intuitive connection 
between the lattice structure and its far-field diffraction pattern 
is possible. A BEC with initially homogeneous phase here accumulates 
a spatially varying phase shift $\Delta \varphi(z) = -V(z)\cdot 
T/\hbar$, and one finds that the far-field momentum distribution is 
$W(p) \propto |\mathrm{FT}(e^{i\Delta\varphi(z)})|^2$, where FT 
denotes the Fourier transform with respect to $z$ \cite{ovchinnikov}. 
For small pulse envelopes, in which the exponential factor can be 
linearized, the time of flight image is directly related to the 
Fourier transform of the potential $V(z)$, i.\,e.\ the reciprocal 
lattice. Within second order we find $W(p) \propto 
\left|\mathrm{FT}\left(1-iV(z)\cdot T/\hbar -(V(z)\cdot 
T/\hbar)^2/2\right)\right|^2$, which allows us to express the 
probability to find an atom in $\pm 1$.\ diffraction orders as $W(\pm 
2\hbar k) \propto \left(\frac{S_1}{4}\right)^2 
\left[1+\left(\frac{S_2}{4}\right)^2\pm\frac{S_2}{2}\sin\Phi\right]$ 
with $S_x = V_x\cdot T/\hbar$.

By choosing variable amplitudes and phases of the lattice potentials, 
different lattices were Fourier synthesized. To begin with, we 
investigated a conventional optical lattice. Fig.~3a shows a typical 
obtained absorption image along with the reconstructed potential 
(which yielded a potential depth $V_1 \simeq {}$3.4\,$\mu$K). For this measurement, only two 
counterpropagating beams with frequency $\omega$ were required. 
Fig.~3b shows corresponding results obtained for the four-photon 
lattice potential ($V_2 \simeq {}$2.5\,$\mu$K), as realized with the 
scheme shown in Fig.~1b. For this measurement, two beams with 
frequency $\omega\pm\Delta\omega$ and one counterpropagating beam 
with frequency $\omega$ were activated. The main peaks of the 
absorption image here are spatially separated by twice the amount 
shown in Fig.~3a, which reflects the by a factor two smaller spatial 
periodicity of $\lambda/4$ of this four-photon lattice.

To demonstrate a Fourier synthesis of atom potentials, two- and 
four-photon lattice potentials were overlapped. Here, all four light 
fields were used (see Fig.~2), so that in addition to the optical 
frequency components shown in Fig.~1b also a counterpropagating beam 
with frequency $\omega$ was active. By changing the phase of the 
control beam with frequency $\omega$ with respect to the other beams, 
the phase of the four-photon contribution could be varied relatively 
to the position of the two-photon potential. Fig.~3c and Fig.~3d show 
far-field diffraction patterns and the reconstructed unit cells for 
such lattices. For the measurement shown in Fig.~3d ($V_1 \simeq 
{}$5.6\,$\mu$K, $V_2 \simeq {}$3.8\,$\mu$K), the phase of the four-photon 
contribution was changed by 180$^\circ$ with respect to that shown in 
Fig.~3c ($V_1 \simeq {}$4.5\,$\mu$K, $V_2 \simeq {}$3.7\,$\mu$K). In both 
images the diffraction pattern is clearly asymmetric, especially when 
considering the amplitude of the $\pm 2\hbar k$ peaks. This is 
attributed as evidence for asymmetric atom potentials, as also 
visible in the shown reconstructed lattice potentials.

We have also investigated diffraction patterns from lattice 
potentials with a more complete range of phase shifts of the 
four-photon potential. As the fundamental period of the combined 
lattice is $\lambda/2$, we anticipate that the relative magnitude of 
the $\pm 2\hbar k$ diffraction peaks gives an indication for the 
asymmetry of the lattice potential. Fig.~4 shows the fraction of 
atoms diffracted into the $\pm 2\hbar k$ orders relatively to the 
total number of atoms as a function of the phase shift. The data 
points have been fitted with sinusoidal functions which is in 
accordance with theory for small pulse envelopes. One clearly 
observes the variation of asymmetry of the diffraction pattern as a 
function of the phase between the different Fourier components. The 
deviations of the fitted curves to the theoretical formula given 
above can be attributed to the fact, that were are not completely in 
the Raman-Nath regime.

To conclude, we report on the realization of atom potentials suitable 
for atomic Bose-Einstein condensates with spatial periodicity 
$\lambda/4$ using multiphoton Raman transitions. This has allowed us 
to demonstrate a scheme for the Fourier synthesis of asymmetric 
shaped lattice potentials. In future, it would be interesting to 
extend this scheme to include even smaller period lattice potentials. 
Besides in the search for new quantum phases, tailored 
high-periodicity lattices are of interest e.\,g.\ in the context of 
subhealing length physics, atom lithography \cite{meschede} and 
quantum computing \cite{jaksch}. An intriguing other perspective 
includes studies of quantum ratchets with atomic quantum gases. It is 
expected that the investigation of the extremely weakly damped 
regime, in which the atomic inertia plays a large role, will allow 
for novel quantum dynamical phenomena \cite{reimann}.

\begin{acknowledgments}
We thank P.~H\"anggi, D.~K\"o\-lle, S.~Flach, and L.~Santos for 
discussions. We acknowledge support by the Deu\-tsche 
For\-schungs\-ge\-mein\-scha\-ft, the Lan\-des\-stif\-tung 
Ba\-den-W\"urttem\-berg and the European Community.
\end{acknowledgments}


\newpage

\section*{Figures}

\begin{figure}[ht]
\includegraphics{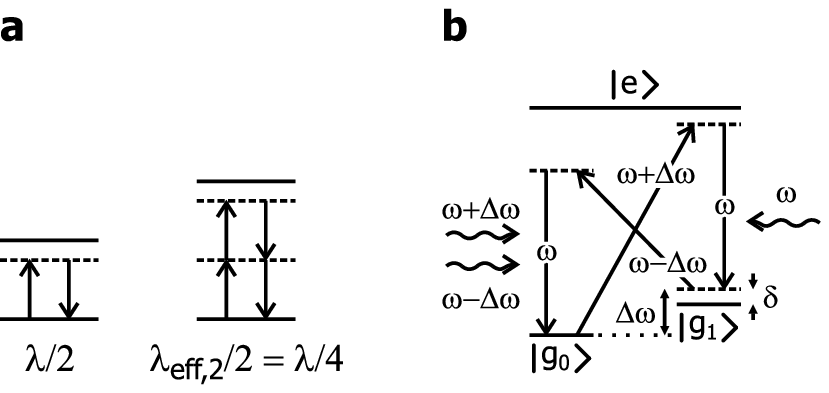}
\caption{(a) Left: Virtual two-photon process in a conventional 
optical lattice, yielding a $\lambda/2$ spatial periodicity of the 
lattice potential. Right: Virtual four-photon process contributing to 
a lattice potential with $\lambda/4$ spatial periodicity. In these 
simple scheme however the usual lattice potential of $\lambda/2$ 
periodicity due to second order processes dominates. (b) Improved 
scheme for generation of a four-photon lattice with $\lambda/4$ 
spatial periodicity, as used in our experimental work. Compared to 
the four-photon process of (a), no second order standing wave 
processes occur. The scheme can be generalized to lattice potentials 
with $\lambda/2n$ periodicity, where $n$ is an integer number.}
\end{figure}

\begin{figure}[ht]
\includegraphics{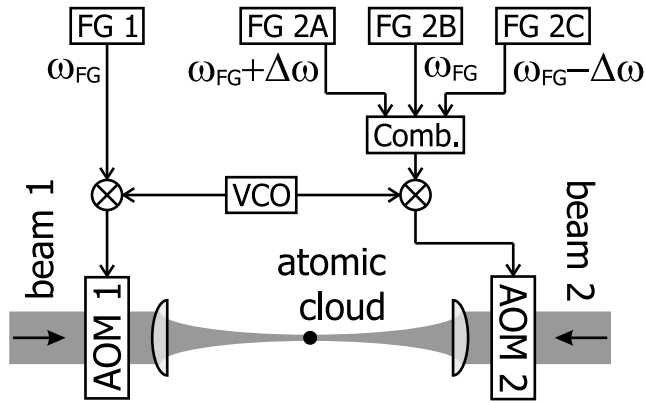}
\caption{Scheme of experimental setup for a Fourier synthesis of 
periodic atom potentials (FG: frequency doubled function generator, 
VCO: voltage controlled oscillator, AOM: acousto-optical modulator). 
Also shown are the frequencies emitted by the rf sources FG1, FG2A, 
FG2B and FG2C, which are used to derive different optical frequency 
components in the lattice beams from a single optical source.}
\end{figure}

\begin{figure}[ht]
\includegraphics{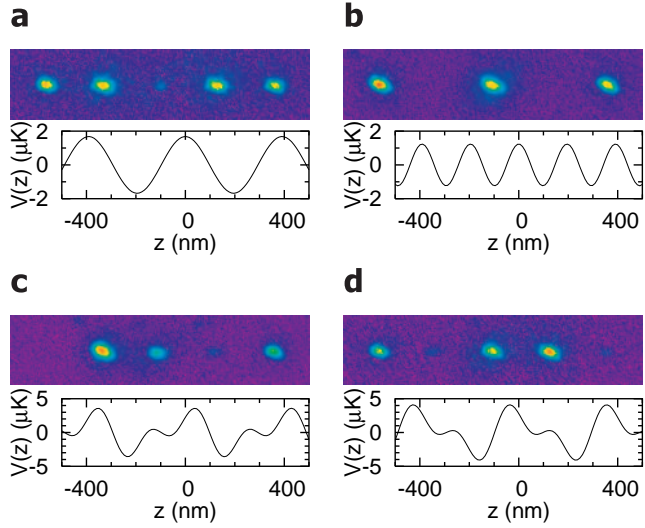}
\caption{Far-field diffraction images of lattice potentials and 
corresponding reconstructed spatial structure of the lattice 
potentials. (a) Two-photon lattice with $\lambda/2$ spatial 
periodicity. (b) Four-photon lattice with $\lambda/4$ spatial 
periodicity. Due to its smaller spatial periodicity, the splitting of 
the clouds is a factor two above that observed in (a). (c) Asymmetric 
lattice realized by superimposing two- and four-photon lattice 
potentials. (d) Same as in (c), but with an additional phase shift of 
180$^\circ$ for the four-photon lattice potential.}
\end{figure}

\begin{figure}[ht]
\includegraphics{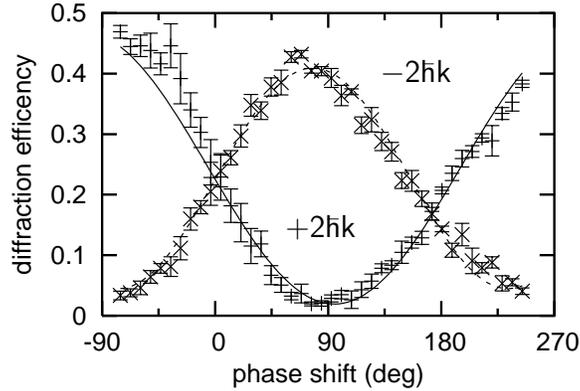}
\caption{Diffraction efficiency into the orders with momentum $+2\hbar 
k$ (bars) and $-2\hbar k$ (crosses) as a function of the phase shift 
of the four-photon component of the lattice potential. Each point 
corresponds to the average of 4 measurements. The data has been 
fitted with sinusoidal functions.}
\end{figure}

\end{document}